\documentclass[12pt]{article}  
\usepackage{graphicx}
\usepackage[a4paper,textwidth=490.0pt,textheight=703.1pt]{geometry}
\usepackage{color}
\usepackage{cite}
\usepackage[rflt]{floatflt}

\usepackage{array}

\usepackage[english]{babel}
\usepackage[latin1]{inputenc}
\usepackage[T1]{fontenc}
\usepackage{ae}

\usepackage{url}

\usepackage{amsmath, amsthm, amssymb}
\newtheorem{thm}{Theorem}[section]

\newtheorem{definition}[thm]{Definition}

\newcommand{\ep}{\varepsilon}

\usepackage{rotating}

\usepackage{algorithm}
\usepackage{algorithmicx}
\usepackage{algpseudocode}
\usepackage{graphicx}

\newcounter{mmacnt}
\def\restartmma{\setcounter{mmacnt}{0}}
\restartmma \catcode`|=\active
\def|#1|{\mathrm{#1}}
\catcode`|=12
\newenvironment{mma}{
 \par\smallskip
 \catcode`|=\active
 \parskip=0pt\parindent=0pt 
 \small
 \def\In##1\\{%
   \def\linebreak{\hfill\break\null\qquad}%
   \refstepcounter{mmacnt}
   \hangindent=2.5em\hangafter=0
   \leavevmode
   \llap{\tiny\sffamily In[\arabic{mmacnt}]:=\kern.5em}%
   \mathversion{bold}\footnotesize$\displaystyle##1$\normalsize
   \mathversion{normal}\par
 }%
 \def\Print##1\\{%
   \def\linebreak{\hfill\break}%
   \hangindent=2.5em\hangafter=0
   \leavevmode ##1\par}%
 \def\Out##1\\{%
   \def\linebreak{$\hfill\break\null\hfill$}%
   \kern\abovedisplayskip\par
   \hangindent=2.5em\hangafter=0
   \leavevmode
   \llap{\tiny\sffamily Out[\arabic{mmacnt}]=\kern.5em}
   \footnotesize$\displaystyle##1$\normalsize\hfill\null\par
   \kern\belowdisplayskip
 }%
 \def\Warning##1##2\\{%
   \def\linebreak{\hfill\break}%
   \hangindent=2.5em\hangafter=0
   \leavevmode
   {\scriptsize##1 : ##2}\par}%
}{%
 \par\smallskip
}

\usepackage{color}

\newenvironment{fshaded}{%
\MakeFramed {\FrameRestore}
}%
{\endMakeFramed}

\usepackage{tikz}
\usetikzlibrary{matrix}

\allowdisplaybreaks[4]

\usepackage{hyperref}

\begin{document}
\setlength{\baselineskip}{0.515cm}
\sloppy
\thispagestyle{empty}
\begin{flushleft}
DESY 20--199
\hfill 
\\
DO-TH 20/06 \\
SAGEX-20-08 \\
January 2021\\
\end{flushleft}

\mbox{}
\vspace*{\fill}
\begin{center}

{\Large\bf The 6th Post-Newtonian Potential Terms at \boldmath $O(G_N^4)$}

\vspace{4cm}
\large
J.~Bl\"umlein$^a$, A.~Maier$^a$, P.~Marquard$^a$, and G.~Sch\"afer$^b$ 

\vspace*{3mm}
{\it  $^a$ Deutsches Elektronen-Synchrotron, DESY,}\\
{\it  Platanenallee 6, D-15738 Zeuthen, Germany}\\

\vspace*{3mm}
{\it $^b$Theoretisch-Physikalisches Institut, Friedrich Schiller-Universit\"at, \\
Max Wien-Platz 1, D--07743 Jena, Germany}\\


\end{center}
\normalsize
\vspace{\fill}
\begin{abstract}
\noindent 
We calculate the potential contributions of the Hamiltonian in harmonic coordinates up 6PN for binary 
mass systems to $O(G_N^4)$ and perform comparisons to recent results in the literature \cite{Bern:2021dqo} 
and \cite{Bini:2020nsb}.
\end{abstract}

\vspace*{\fill}
\noindent
\newpage

\noindent
Gravitational wave signals from merging black holes and neutron stars \cite{LIGO,PROJECT} provide tests 
of precision predictions within Einstein gravity for the dynamics of binary systems of massive objects. 
Their dynamics is calculated using  post-Newtonian 
\cite{FOURTH,Blumlein:2020pog,Blumlein:2019zku,FIFTH,Blumlein:2020pyo,Blumlein:2020znm,Bini:2020nsb,Bini:2020hmy} 
and the post-Minkowskian \cite{POSTMINK,Bern:2021dqo}\footnote{See also Ref.~[12] in 
\cite{Blumlein:2020znm}.} methods\footnote{For surveys see \cite{SURVEYS}.}. To match present and future measurements, 
high order computations are necessary. In the post-Newtonian (PN) case the level of 6PN 
\cite{Blumlein:2020znm,Bini:2020nsb,Bini:2020hmy} has now been reached for the conservative part of the two-body motion,
while in the post-Minkowskian approach the calculation of the $O(G_N^4)$ potential terms is the most
far reaching result \cite{Bern:2021dqo}. Here $G_N$ denotes Newton's constant.
The different calculations are often performed using different 
gauges (harmonic, ADM, isotropic and EOB) in deriving the Lagrangian or Hamiltonian. It is important to 
cross check the results between the two methods, and to compare different representations in the most 
general way possible. One either can perform special comparisons in calculating the same observable
or one performs a canonical transformation \cite{CAN} between the different Hamiltonians obtained. The 
latter result implies that the results for all observables are the same because of the invariance of the
action.

In this paper we report about the calculation of the $O(G_N^4)$ potential terms to 6PN, extending 
previous work covering the terms up to $O(G_N^3)$. We use the effective field theory approach 
\cite{METH}, for details see \cite{Blumlein:2020znm}. We work in the harmonic gauge and $D = 4 
- 2\ep$ dimensions. 
The Feynman diagrams are generated using {\tt QGRAF} \cite{Nogueira:1991ex}. The Lorentz algebra is 
carried out using {\tt Form} \cite{FORM} and we perform the integration by parts (IBP) reduction to 
master integrals using the code {\tt Crusher} \cite{CRUSHER}. 
Table~\ref{TAB1} gives an overview on the present calculation.
\begin{table}[H]\centering
\begin{tabular}{rrrrrrr}
\hline
\multicolumn{1}{c}{\#loops }            &
\multicolumn{1}{c}{QGRAF}               &
\multicolumn{1}{c}{source irred}        &
\multicolumn{1}{c}{no source loops}     & 
\multicolumn{1}{c}{no tadpoles}         &
\multicolumn{1}{c}{symmetrised}         &
\multicolumn{1}{c}{masters}             
\\
\hline
  0 & 3      & 3     & 3     & 3     & 3      & \\
  1 & 72     & 72    & 72    & 72    & 24     & 1 \\ 
  2 & 4322   & 4322  & 4322  & 3512  & 485    & 1  \\
  3 & 111752 & 86900 & 85467 & 61863 & 5553   & 2 \\
\hline
\end{tabular}
\caption[]{\sf Numbers of contributing diagrams at the different loop levels and master integrals.}
\label{TAB1}
\end{table}

\noindent
Redundant diagrams are eliminated in a series of steps outlined in Ref.~\cite{Blumlein:2019zku}.
6065 diagrams do finally contribute to the present result. The computation time amounted to a few days on 
an {\tt Intel(R) Xeon(R) CPU E5-2643 v4}. The Lagrange function of $m$th order, still  containing the 
accelerations $a_i$ and time derivatives thereof, are converted into a first order Lagrange density by 
applying double zero insertions \cite{Schafer:1984mr,Damour:1985mt} together with partial integration and 
the remaining linear accelerations by a shift \cite{SHIFT,Damour:1985mt}, cf.~\cite{Blumlein:2020pog}.
By this operation we leave harmonic coordinates. A Legendre transformation leads then to the
potential contributions of the  Hamiltonian, which still contains pole terms in the dimensional parameter 
$\ep$.

We define $H = (H_{\rm real} - M c^2)/(\mu c^2)$ with $M = m_1 + m_2, \mu = m_1 m_2/M, \nu = \mu/M$, with 
$m_{1,2}$ the two masses of the binary system, $c$ the velocity of light, and obtain the following result 
for the potential contributions\footnote{including the kinetic terms.} 
to the 6PN Hamiltonian to $O(G_N^4)$
\begin{eqnarray}
\label{eq:HPOT65}
H_{\rm pot} &=&
\frac{p^2}{2}
+\frac{1}{8} p^4 (-1+3 \nu )
+\frac{1}{16} p^6 \Biggl(
        1-5 \nu +5 \nu ^2\Biggr)
+p^8 \Biggl(
        -\frac{5}{128}+\frac{35 \nu }{128}-\frac{35 \nu ^2}{64}+\frac{35 \nu ^3}{128}\Biggr)
\nonumber\\ &&
+p^{10} \Biggl(
        \frac{7}{256}-\frac{63 \nu }{256}+\frac{189 \nu ^2}{256}-\frac{105 \nu ^3}{128}+\frac{63 \nu
^4}{256}\Biggr)
+p^{12} \Biggl(
        -\frac{21}{1024}+\frac{231 \nu }{1024}-\frac{231 \nu ^2}{256}
\nonumber\\ &&
+\frac{1617 \nu ^3}{1024}-\frac{1155
\nu ^4}{1024}+\frac{231 \nu ^5}{1024}\Biggr)
+p^{14} \Biggl(
        \frac{33}{2048}-\frac{429 \nu }{2048}+\frac{2145 \nu ^2}{2048}-\frac{1287 \nu ^3}{512}
\nonumber\\ &&
+\frac{3003
\nu ^4}{1024}-\frac{3003 \nu ^5}{2048}+\frac{429 \nu ^6}{2048}\Biggr)
\nonumber\\
&&
+\frac{1}{r} \Biggl[
        -1
        +\frac{1}{2} p^2 (-3-\nu )
        -\frac{(p.n)^2 \nu }{2}
        -\frac{1}{4} p^2 (p.n)^2 (-1+\nu ) \nu
        -\frac{3}{8} (p.n)^4 \nu ^2
        \nonumber\\
        &&        -\frac{3}{16} p^2 (p.n)^4 (-1+\nu ) \nu ^2
-\frac{5}{16} (p.n)^6 \nu ^3
        -\frac{5}{32} p^2 (p.n)^6 (-2+\nu ) \nu ^3
        -\frac{35}{128} (p.n)^8 \nu ^4
        \nonumber\\
        &&        -\frac{63}{256} (p.n)^{10} \nu ^5
-\frac{231 (p.n)^{12} \nu ^6}{1024}
        +\frac{1}{8} p^4 \Biggl(
                5-22 \nu -3 \nu ^2\Biggr)
        +p^6 \Biggl(
                -\frac{7}{16}+\frac{45 \nu }{16}
\nonumber\\ &&
-\frac{31 \nu ^2}{8}-\frac{5 \nu ^3}{16}\Biggr)
        +p^4 (p.n)^2 \Biggl(
                -\frac{3 \nu }{16}+\frac{11 \nu ^2}{16}-\frac{3 \nu ^3}{16}\Biggr)
        +p^8 \Biggl(
                \frac{45}{128}-\frac{95 \nu }{32}+\frac{475 \nu ^2}{64}-\frac{267 \nu ^3}{64}
\nonumber\\ &&
-\frac{35 \nu
^4}{128}\Biggr)
        +p^6 (p.n)
        ^2 \Biggl(
                \frac{5 \nu }{32}-\frac{29 \nu ^2}{32}+\frac{11 \nu ^3}{32}-\frac{5 \nu ^4}{32}\Biggr)
        +p^4 (p.n)^4 \Biggl(
                -\frac{9 \nu ^2}{64}+\frac{33 \nu ^3}{64}-\frac{9 \nu ^4}{64}\Biggr)
        \nonumber\\
        &&+p^{10} \Biggl(
                -\frac{77}{256}+\frac{805 \nu }{256}-\frac{2865 \nu ^2}{256}+\frac{3995 \nu
^3}{256}-\frac{1615 \nu ^4}{256}-\frac{63 \nu ^5}{256}\Biggr)
        \nonumber\\
        &&+p^2 (p.n)^8 \Biggl(
                -\frac{35 \nu ^3}{256}+\frac{105 \nu ^4}{128}-\frac{35 \nu ^5}{256}\Biggr)
        +p^8 (p.n)^2 \Biggl(
                -\frac{35 \nu }{256}+\frac{275 \nu ^2}{256}-\frac{221 \nu ^3}{64}
+\frac{867 \nu
^4}{256}
        \nonumber\\
        &&-\frac{35 \nu ^5}{256}\Biggr)
+p^4 (p.n)^6 \Biggl(
                -\frac{15 \nu ^3}{128}-\frac{45 \nu ^4}{128}-\frac{15 \nu ^5}{128}\Biggr)
        +p^6 (p.n)^4 \Biggl(
                \frac{15 \nu ^2}{128}-\frac{3 \nu ^3}{128}-\frac{3 \nu ^4}{64}-\frac{15 \nu ^5}{128}\Biggr)
        \nonumber\\
        &&+p^2 (p.n)^{10} \Biggl(
                \frac{63 \nu ^3}{512}-\frac{441 \nu ^4}{512}+\frac{945 \nu ^5}{512}-\frac{63 \nu
^6}{512}\Biggr)
        +p^4 (p.n)^8 \Biggl(
                -\frac{105 \nu ^3}{512}+
                \frac{1505 \nu ^4}{1024}-\frac{1785 \nu ^5}{512}
        \nonumber\\
        &&-\frac{105 \nu ^6}{1024}\Biggr)
+p^8 (p.n)^4 \Biggl(
                -\frac{105 \nu ^2}{1024}+\frac{2169 \nu ^3}{1024}-\frac{6891 \nu ^4}{1024}+\frac{3003 \nu
^5}{1024}-\frac{105 \nu ^6}{1024}\Biggr)
        \nonumber\\
        &&+p^6 (p.n)^6 \Biggl(
                -\frac{5 \nu ^3}{64}-\frac{5 \nu ^4}{8}+\frac{1025 \nu ^5}{256}-\frac{25 \nu ^6}{256}\Biggr)
        \nonumber\\
        &&+p^{12} \Biggl(
                \frac{273}{1024}-\frac{1701 \nu }{512}+\frac{15631 \nu ^2}{1024}-\frac{30277 \nu
^3}{1024}+\frac{5007 \nu ^4}{256}+\frac{597 \nu ^5}{1024}-\frac{231 \nu ^6}{1024}\Biggr)
        \nonumber\\
        &&+p^{10} (p.n)^2 \Biggl(
                \frac{63 \nu }{512}-\frac{623 \nu ^2}{512}+\frac{87 \nu ^3}{64}+\frac{95 \nu
^4}{16}-\frac{1889 \nu ^5}{256}-\frac{63 \nu ^6}{512}\Biggr)
\Biggr]
\nonumber\\ &&
+\frac{1}{r^2} \Biggl[
        \frac{1}{2}
        +\frac{1}{2} (p.n)^2 (-1+6 \nu )
        +\frac{1}{16} p^4 (-1+8 \nu ) (29+12 \nu )
        +\frac{1}{4} p^2 (11+15 \nu )
\nonumber\\ &&
  +\frac{1}{3} (p.n)^4 \nu  (7+69 \nu )
        +\frac{1}{4} p^2 (p.n)^2 \Biggl(
                1-36 \nu -36 \nu ^2\Biggr)
        +p^2 (p.n)^4 \Biggl(
                -
                \frac{79 \nu }{192}-\frac{4737 \nu ^2}{64}
\nonumber\\ &&
-\frac{7511 \nu ^3}{96}\Biggr)
        +p^6 \Biggl(
                \frac{55}{32}-\frac{667 \nu }{64}+\frac{1217 \nu ^2}{64}-\frac{89 \nu ^3}{64}\Biggr)
        +p^4 (p.n)^2 \Biggl(
                -\frac{3}{16}-\frac{99 \nu }{16}+\frac{733 \nu ^2}{16}
        \nonumber\\ &&
+\frac{3189 \nu ^3}{64}\Biggr)
        +(p.n)^6 \Biggl(
                \frac{487 \nu }{160}+\frac{543 \nu ^2}{32}+\frac{4609 \nu ^3}{80}\Biggr)
        +p^2 (p.n)^6 \Biggl(
                -\frac{5117 \nu }{320}-\frac{8331 \nu ^2}{160}
\nonumber\\ &&
+\frac{136977 \nu ^3}{1280}-\frac{1178329 \nu
^4}{1280}\Biggr)
        +p^6 (p.n)^2 \Biggl(
                \frac{5}{32}-\frac{589 \nu }{16}+\frac{4969 \nu ^2}{64}+\frac{177 \nu ^3}{256}-\frac{62143
\nu ^4}{256}\Biggr)
        \nonumber\\
        &&+p^8 \Biggl(
                -\frac{445}{256}+\frac{937 \nu }{32}-\frac{11535 \nu ^2}{128}+\frac{16283 \nu
^3}{256}+\frac{6649 \nu ^4}{256}\Biggr)
+(p.n)^8 \Biggl(
                \frac{159 \nu }{28}+\frac{751 \nu ^2}{28}
        \nonumber\\
        &&
-\frac{289839 \nu ^3}{4480}+\frac{1443091 \nu
^4}{4480}\Biggr)
+p^4 (p.n)^4 \Biggl(
                \frac{8951 \nu }{384}+
                \frac{925 \nu ^2}{24}-\frac{125225 \nu ^3}{768}+\frac{652381 \nu ^4}{768}\Biggr)
        \nonumber\\
        &&+p^4 (p.n)^6 \Biggl(
                \frac{9647 \nu }{320}+\frac{42947 \nu ^2}{320}-\frac{4799 \nu ^3}{8}+\frac{4064307 \nu
^4}{2560}-\frac{98191 \nu ^5}{80}\Biggr)
        \nonumber\\
        &&+p^8 (p.n)^2 \Biggl(
                -\frac{35}{256}-\frac{6747 \nu }{128}+\frac{14485 \nu ^2}{32}-\frac{1169335 \nu
^3}{1024}+\frac{1033543 \nu ^4}{1024}-\frac{289599 \nu ^5}{1024}\Biggr)
        \nonumber\\
        &&+(p.n)^{10} \Biggl(
                \frac{5333 \nu }{1152}+\frac{40343 \nu ^2}{1152}-\frac{1815535 \nu ^3}{16128}+\frac{2108389
\nu ^4}{8064}+\frac{87329 \nu ^5}{16128}\Biggr)
        \nonumber\\
        &&+p^{10} \Biggl(
                \frac{917}{512}+\frac{315 \nu }{256}-\frac{26259 \nu ^2}{512}+\frac{125919 \nu
^3}{1024}-\frac{87953 \nu ^4}{1024}+\frac{16495 \nu ^5}{1024}\Biggr)
        \nonumber\\
        &&+p^2 (p.n)^8 \Biggl(
                -\frac{25007 \nu }{1792}-\frac{1226481 \nu ^2}{8960}+\frac{7963357 \nu
^3}{17920}-\frac{9471627 \nu ^4}{8960}+\frac{7130709 \nu ^5}{17920}\Biggr)
        \nonumber\\
        &&+p^6 (p.n)^4 \Biggl(
                -\frac{7201 \nu }{1536}
                -
                \frac{222197 \nu ^2}{768}+\frac{53039 \nu ^3}{48}-\frac{1277989 \nu ^4}{768}+\frac{390901
\nu ^5}{384}\Biggr)
\Biggr]
\nonumber\\
&&+\frac{1}{r^3} \Biggl[
        -\frac{1}{2}
        -\frac{\nu }{4}
        +p^2 \Biggl(
                -\frac{17}{4}+\frac{643 \nu }{72}-\frac{7 \pi ^2 \nu }{8}-\frac{3 \nu ^2}{2}\Biggr)
        +(p.n)^2 \Biggl(
                \frac{3}{2}-\frac{1013 \nu }{12}+\frac{21 \pi ^2 \nu }{8}
        \nonumber\\
        &&+\frac{49 \nu ^2}{4}\Biggr)
+(p.n)^4 \Biggl(
                -\frac{6695 \nu }{32}+\frac{4395 \pi ^2 \nu }{1024}-
                \frac{200369 \nu ^2}{320}+\frac{345 \pi ^2 \nu ^2}{128}
-\frac{333 \nu ^3}{32}\Biggr)
\nonumber\\ &&
        +p^2 (p.n)^2 \Biggl(
                -\frac{5}{4}+\frac{294477 \nu }{800}-\frac{2955 \pi ^2 \nu }{512}+\frac{167173 \nu
^2}{1200}+\frac{1095 \pi ^2 \nu ^2}{128}+\frac{11 \nu ^3}{16}\Biggr)
        \nonumber\\
        &&+p^4 \Biggl(
                \frac{65}{16}-\frac{94439 \nu }{800}+\frac{1091 \pi ^2 \nu }{1024}+\frac{319789 \nu
^2}{14400}-\frac{217 \pi ^2 \nu ^2}{64}+\frac{205 \nu ^3}{32}\Biggr)
+(p.n)^6 \Biggl(
                -\frac{627281 \nu }{960}
        \nonumber\\
        &&
+\frac{42105 \pi ^2 \nu }{4096}+\frac{18031 \nu
^2}{3360}+\frac{14175 \pi ^2 \nu ^2}{4096}-\frac{14830647 \nu ^3}{4480}-\frac{65625 \pi ^2 \nu
^3}{1024}-\frac{17623 \nu ^4}{240}\Biggr)
        \nonumber\\
        &&+p^2 (p.n)^4 \Biggl(
                \frac{31715507 \nu }{23520}-\frac{89625 \pi ^2 \nu }{4096}+\frac{848889 \nu
^2}{1568}+\frac{127125 \pi ^2 \nu ^2}{4096}+\frac{373945981 \nu ^3}{94080}
        \nonumber\\
&& -\frac{30075 \pi ^2 \nu
^3}{1024}-\frac{5749 \nu ^4}{96}\Biggr)
        +p^6 \Biggl(
                -\frac{161}{32}+\frac{11206267 \nu }{141120}-\frac{7719 \pi ^2 \nu }{4096}+
                \frac{3605263 \nu ^2}{29400}
\nonumber\\ &&
+\frac{29987 \pi ^2 \nu ^2}{4096}+\frac{108551131 \nu
^3}{4233600}-\frac{20259 \pi ^2 \nu ^3}{1024}-\frac{593 \nu ^4}{32}\Biggr)
+p^4 (p.n)^2 \Biggl(
                \frac{21}{16}-\frac{162949463 \nu }{235200}
\nonumber\\ &&
+\frac{58887 \pi ^2 \nu }{4096}-\frac{1945067
\nu ^2}{2450}-\frac{172311 \pi ^2 \nu ^2}{4096}-\frac{2369976949 \nu ^3}{1411200}+\frac{106947 \pi ^2 \nu
^3}{1024}+\frac{549 \nu ^4}{32}\Biggr)
        \nonumber\\
        &&+p^4 (p.n)^4 \Biggl(
                -\frac{132847139 \nu }{141120}+\frac{3302175 \pi ^2 \nu }{65536}-\frac{22546873057 \nu
^2}{1693440}-\frac{413055 \pi ^2 \nu ^2}{4096}
\nonumber\\ &&
+\frac{17629672339 \nu ^3}{1128960}+\frac{1825635 \pi ^2 \nu
^3}{4096}-\frac{47772068147 \nu ^4}{846720}-\frac{12653055 \pi ^2 \nu ^4}{8192}-\frac{1431581 \nu
^5}{768}\Biggr)
        \nonumber\\
        &&+(p.n)^8 \Biggl(
                \frac{4756417 \nu }{26880}+\frac{2925405 \pi ^2 \nu }{131072}-\frac{192771863 \nu
^2}{40320}+\frac{23625 \pi ^2 \nu ^2}{4096}+\frac{23044829 \nu ^3}{10752}
\nonumber\\ &&
-\frac{784665 \pi ^2 \nu
^3}{4096}-\frac{1282851793 \nu ^4}{32256}-
                \frac{9135 \pi ^2 \nu ^4}{32}-\frac{1816323 \nu ^5}{2560}\Biggr)
+p^8 \Biggl(
                \frac{1605}{256}-\frac{18459883 \nu }{268800}
        \nonumber\\
        &&
+\frac{310029 \pi ^2 \nu
}{131072}
+\frac{1275787309 \nu ^2}{3175200}-\frac{23771 \pi ^2 \nu ^2}{2048}-\frac{38805398273 \nu
^3}{25401600}+\frac{73075 \pi ^2 \nu ^3}{2048}
\nonumber\\ &&
+\frac{174300143 \nu ^4}{268800}-\frac{446081 \pi ^2 \nu
^4}{8192}-\frac{7797 \nu ^5}{512}\Biggr)
+p^6 (p.n)^2 \Biggl(
                -\frac{45}{32}-\frac{204352217 \nu }{2822400}
        \nonumber\\
        &&
-\frac{685233 \pi ^2 \nu
}{32768}+\frac{11064793483 \nu ^2}{2116800}+\frac{291459 \pi ^2 \nu ^2}{4096}-\frac{6068679041 \nu
^3}{1693440}-\frac{1182561 \pi ^2 \nu ^3}{4096}
\nonumber\\ &&
+\frac{26379298087 \nu ^4}{2822400}+\frac{5051661 \pi ^2 \nu
^4}{8192}+\frac{54337 \nu ^5}{128}\Biggr)
+p^2 (p.n)^6 \Biggl(
                \frac{257519 \nu }{336}-\frac{1824025 \pi ^2 \nu }{32768}
        \nonumber\\
        &&
+\frac{8649135391 \nu
^2}{725760}
+\frac{212625 \pi ^2 \nu ^2}{4096}-\frac{81220205 \nu ^3}{8064}-\frac{209335 \pi ^2 \nu
^3}{4096}+\frac{8667605527 \nu ^4}{103680}
\nonumber\\ &&
+\frac{10868515 \pi ^2 \nu ^4}{8192}+
                \frac{1100997 \nu ^5}{640}\Biggr)
\Biggr]
\nonumber\\
&&+\frac{1}{r^4} \Biggl[
        \frac{3}{8}
        -\frac{1279 \nu }{72}
        +\frac{15 \pi ^2 \nu }{64}
        +(p.n)^2 \Biggl(
                -\frac{11}{4}+\frac{73801 \nu }{1600}-\frac{4429 \pi ^2 \nu }{192}+\frac{953891 \nu
^2}{7200}
\nonumber\\ &&
-\frac{4033 \pi ^2 \nu ^2}{128}\Biggr)
        +p^2 \Biggl(
                \frac{95}{16}+\frac{115733 \nu }{2880}+\frac{643 \pi ^2 \nu }{128}-\frac{1223723 \nu
^2}{7200}+\frac{1419 \pi ^2 \nu ^2}{128}\Biggr)
        \nonumber\\
        &&+(p.n)^4 \Biggl(
                -\frac{1}{8}+\frac{1895797259 \nu }{235200}-\frac{3293913 \pi ^2 \nu
}{4096}-\frac{1742633989 \nu ^2}{117600}+\frac{2617363 \pi ^2 \nu ^2}{4096}
\nonumber\\ &&
+\frac{14035555739 \nu
^3}{705600}-\frac{361499 \pi ^2 \nu ^3}{1024}\Biggr)
+p^4 \Biggl(
                -\frac{499}{64}+\frac{2128837091 \nu }{1411200}-\frac{1328147 \pi ^2 \nu
}{12288}
\nonumber\\ &&
-\frac{420686323 \nu ^2}{132300}+\frac{2076041 \pi ^2 \nu ^2}{12288}+\frac{617770201 \nu
^3}{423360}+
                \frac{98447 \pi ^2 \nu ^3}{3072}\Biggr)
        +p^2 (p.n)^2 \Biggl(
                \frac{29}{8}
\nonumber\\ &&
-\frac{2385014243 \nu }{282240}+\frac{5042575 \pi ^2 \nu
}{6144}+\frac{35606467999 \nu ^2}{2116800}-\frac{5962205 \pi ^2 \nu ^2}{6144}-\frac{3656476457 \nu
^3}{235200}
        \nonumber\\ &&
+\frac{131231 \pi ^2 \nu ^3}{1536}\Biggr)
        +p^2 (p.n)^4 \Biggl(
                \frac{3}{16}+\frac{175079560811 \nu }{940800}-\frac{93462353 \pi ^2 \nu
}{8192}-\frac{3559323922849 \nu ^2}{2822400}
\nonumber\\ &&
+\frac{166850287 \pi ^2 \nu ^2}{4096}+\frac{1804730974343 \nu
^3}{1881600}+\frac{219605317 \pi ^2 \nu ^3}{8192}+\frac{906978233137 \nu ^4}{1693440}
\nonumber\\ &&
-\frac{631940135 \pi
^2 \nu ^4}{8192}\Biggr)
+p^6 \Biggl(
                \frac{1567}{128}+\frac{331187219953 \nu }{50803200}-\frac{9597775 \pi ^2 \nu
}{24576}-\frac{242295730217 \nu ^2}{8467200}
\nonumber\\ &&
+\frac{343433 \pi ^2 \nu ^2}{1536}-\frac{10130224103 \nu
^3}{10160640}+\frac{72402467 \pi ^2 \nu ^3}{24576}-\frac{30642112157 \nu ^4}{4233600}+\frac{22396811 \pi ^2
\nu ^4}{24576}\Biggr)
        \nonumber\\
        &&+p^4 (p.n)^2 \Biggl(
                -\frac{165}{32}-
                \frac{1488934043759 \nu }{16934400}+\frac{32454227 \pi ^2 \nu }{6144}+\frac{135138293977
\nu ^2}{282240}
\nonumber\\ &&
-\frac{266286257 \pi ^2 \nu ^2}{24576}-\frac{574199075573 \nu ^3}{3386880}-\frac{110076807
\pi ^2 \nu ^3}{4096}-\frac{1067026232959 \nu ^4}{8467200}
\nonumber\\ &&
+\frac{41375245 \pi ^2 \nu ^4}{3072}\Biggr)
+(p.n)^6 \Biggl(
                -\frac{313940389879 \nu }{3175200}+\frac{78229555 \pi ^2 \nu }{12288}+\frac{332644084181
\nu ^2}{403200}
\nonumber\\ &&
-\frac{274433707 \pi ^2 \nu ^2}{8192}-\frac{44775952005941 \nu ^3}{50803200}+\frac{17322247
\pi ^2 \nu ^3}{3072}-\frac{5330869608031 \nu ^4}{12700800}
\nonumber\\ &&
+\frac{846593393 \pi ^2 \nu ^4}{12288}\Biggr)
\Biggr]
\nonumber\\
&&+\frac{1}{\ep} \Biggl[
        \frac{1}{r^3} \Biggl[
                -\frac{17 p^2 \nu }{6}
                +\frac{17 (p.n)^2 \nu }{2}
                +\frac{5}{3} (p.n)^4 \nu  (12+37 \nu )
                -\frac{11}{60} p^2 (p.n)^2 \nu  (195+133 \nu )
\nonumber\\ &&
  -\frac{1}{180} p^4 \nu  (-1425+757 \nu )
                -
                \frac{1}{56} p^2 (p.n)^4 \nu  \Biggl(
                        4536+7280 \nu
+18979 \nu ^2\Biggr) +p^6 \Biggl(
                        \frac{1173 \nu }{80}
\nonumber\\ &&
-\frac{13583 \nu ^2}{336}-\frac{6889 \nu ^3}{360}\Biggr)
                +p^4 (p.n)^2 \Biggl(
                        -\frac{2271 \nu }{80}+\frac{23047 \nu ^2}{112}+\frac{16538 \nu ^3}{105}\Biggr)
                \nonumber\\
                &&+(p.n)^6 \Biggl(
                        77 \nu -\frac{91 \nu ^2}{6}+\frac{2891 \nu ^3}{12}\Biggr)
+p^4 (p.n)^4 \Biggl(
                        \frac{8469 \nu }{56}+\frac{80489 \nu ^2}{168}+\frac{243865 \nu ^3}{672}
                \nonumber\\
                &&-\frac{25805
\nu ^4}{28}\Biggr)
+(p.n)^8 \Biggl(
                        \frac{333 \nu }{2}+106 \nu ^2+290 \nu ^3-\frac{1789 \nu ^4}{2}\Biggr)
+p^8 \Biggl(
                        \frac{23587 \nu }{672}
                \nonumber\\
                &&
-\frac{2158799 \nu ^2}{15120}
+\frac{823993 \nu
^3}{3780}-\frac{758113 \nu ^4}{6048}\Biggr)
                +p^6 (p.n)^2 \Biggl(
                        -\frac{149963 \nu }{1120}+\frac{1677317 \nu ^2}{5040}
\nonumber\\ &&
-\frac{1954571 \nu
^3}{2520}+\frac{607031 \nu ^4}{1008}\Biggr)
                +p^2 (p.n)^6 \Biggl(
                        -\frac{549 \nu }{2}-\frac{9545 \nu ^2}{18}-\frac{129523 \nu ^3}{288}+\frac{419977
\nu ^4}{288}\Biggr)
        \Biggr]
        \nonumber\\
        &&+
        \frac{1}{r^4} \Biggl[
                \frac{17 \nu }{6}
                -\frac{1}{180} p^2 \nu  (-1215+827 \nu )
                +\frac{1}{180} (p.n)^2 \nu  (-3354+10685 \nu )
                +(p.n)^4 \Biggl(
                        -\frac{13059 \nu }{70}
\nonumber\\ &&
+\frac{16223 \nu ^2}{28}-\frac{304669 \nu ^3}{240}\Biggr)
                +p^4 \Biggl(
                        -\frac{49023 \nu }{560}+\frac{592957 \nu ^2}{2520}-\frac{297509 \nu ^3}{2520}\Biggr)
+p^2 (p.n)^2 \Biggl(
                        \frac{25169 \nu }{105}
\nonumber\\ &&
-\frac{1895597 \nu ^2}{2520}+\frac{352834 \nu ^3}{315}\Biggr)
                +(p.n)^6 \Biggl(
                        \frac{952841 \nu }{630}-\frac{7199557 \nu ^2}{210}+\frac{76415687 \nu
^3}{1260}
                \nonumber\\
                &&-\frac{214306523 \nu ^4}{10080}\Biggr)
+p^4 (p.n)^2 \Biggl(
                        \frac{1324003 \nu }{560}-\frac{98849389 \nu ^2}{3360}+\frac{102404527 \nu
^3}{3360}
\nonumber\\ &&
-\frac{2476553 \nu ^4}{1120}\Biggr)
                +p^6 \Biggl(
                        -\frac{652717 \nu }{2016}+\frac{19320373 \nu ^2}{7560}-\frac{19664101 \nu
^3}{15120}-\frac{18212479 \nu ^4}{30240}\Biggr)
                \nonumber\\
                &&+p^2 (p.n)^4 \Biggl(
                        -\frac{448223 \nu }{105}
                        +
                        \frac{13413467 \nu ^2}{210}-\frac{155079209 \nu ^3}{1680}+\frac{10165237 \nu
^4}{420}\Biggr)
        \Biggr]
\Biggr]
\nonumber\\
&& + \ln\Biggl(
        \frac{r}{r_0} \Biggr)
\Biggl[
        \frac{1}{r^3} \Biggl[
                -17 p^2 \nu
                +51 (p.n)^2 \nu
                +10 (p.n)^4 \nu  (12+37 \nu )
                -\frac{11}{10} p^2 (p.n)^2 \nu  (195+133 \nu )
                \nonumber\\
                &&-\frac{1}{30} p^4 \nu  (-1425+757 \nu )
                +\frac{7}{2} (p.n)^6 \nu  \Biggl(
                        132-26 \nu +413 \nu ^2\Biggr)
                -\frac{3}{28} p^2 (p.n)^4 \nu  \Biggl(
                        4536+7280 \nu
                \nonumber\\ &&
+18979 \nu ^2\Biggr)
                +p^6 \Biggl(
                        \frac{3519 \nu }{40}-\frac{13583 \nu ^2}{56}-\frac{6889 \nu ^3}{60}\Biggr)
                +p^4 (p.n)^2 \Biggl(
                        -\frac{6813 \nu }{40}+\frac{69141 \nu ^2}{56}
\nonumber\\ &&
+\frac{33076 \nu ^3}{35}\Biggr)
                +p^4 (p.n)^4 \Biggl(
                        \frac{25407 \nu }{28}+\frac{80489 \nu ^2}{28}+\frac{243865 \nu ^3}{112}-\frac{77415
\nu ^4}{14}\Biggr)
                \nonumber\\
                &&+(p.n)^8 \Biggl(
                        999 \nu +636 \nu ^2+1740 \nu ^3-5367 \nu ^4\Biggr)
                +p^8 \Biggl(
                        \frac{23587 \nu }{112}-
                        \frac{2158799 \nu ^2}{2520}+\frac{823993 \nu ^3}{630}
\nonumber\\ &&
-\frac{758113 \nu
^4}{1008}\Biggr)
                +p^6 (p.n)^2 \Biggl(
                        -\frac{449889 \nu }{560}+\frac{1677317 \nu ^2}{840}-\frac{1954571 \nu
^3}{420}+\frac{607031 \nu ^4}{168}\Biggr)
                \nonumber\\
                &&+p^2 (p.n)^6 \Biggl(
                        -1647 \nu -\frac{9545 \nu ^2}{3}-\frac{129523 \nu ^3}{48}+\frac{419977 \nu
^4}{48}\Biggr)
        \Biggr]
        \nonumber\\
        &&+\frac{1}{r^4} \Biggl[
                \frac{68 \nu }{3}
                -\frac{2}{45} p^2 \nu  (-1215+827 \nu )
                +\frac{2}{45} (p.n)^2 \nu  (-3354+10685 \nu )
                +(p.n)^4 \Biggl(
                        -\frac{52236 \nu }{35}
                \nonumber\\ &&
+\frac{32446 \nu ^2}{7}-\frac{304669 \nu ^3}{30}\Biggr)
                +p^4 \Biggl(
                        -\frac{49023 \nu }{70}+\frac{592957 \nu ^2}{315}-\frac{297509 \nu ^3}{315}\Biggr)
\nonumber\\ &&
+p^2 (p.n)^2 \Biggl(
                        \frac{201352 \nu }{105}-\frac{1895597 \nu ^2}{315}+\frac{2822672 \nu ^3}{315}\Biggr)
                +(p.n)^6 \Biggl(
                        \frac{3811364 \nu }{315}-\frac{28798228 \nu ^2}{105}
\nonumber\\ &&
+\frac{152831374 \nu
^3}{315}-\frac{214306523 \nu ^4}{1260}\Biggr)
                +p^4 (p.n)^2 \Biggl(
                        \frac{1324003 \nu }{70}-\frac{98849389 \nu ^2}{420}+\frac{102404527 \nu
^3}{420}
\nonumber\\ &&
-\frac{2476553 \nu ^4}{140}\Biggr)
                +p^6 \Biggl(
                        -\frac{652717 \nu }{252}+\frac{19320373 \nu ^2}{945}-\frac{19664101 \nu
^3}{1890}-\frac{18212479 \nu ^4}{3780}\Biggr)
                \nonumber\\
                &&+p^2 (p.n)^4 \Biggl(
                        -\frac{3585784 \nu }{105}+\frac{53653868 \nu ^2}{105}-\frac{155079209 \nu
^3}{210}+\frac{20330474 \nu ^4}{105}\Biggr)
        \Biggr]
        \Biggr],
\end{eqnarray}
 
working in cms coordinates and using the rescaling defined in \cite{Blumlein:2020pyo}, Eq.~(7). In 
dot-products the vectors are 3--vectors. Otherwise the same symbol denotes their modulus and
$p.n = p.r/r$. The implicit counting of the powers in $\eta^2 \equiv 1/c^2$ is defined 
in \cite{Blumlein:2020pog}, Eq.~(54).

In the limit $\nu \rightarrow 0$ we agree with the Schwarzschild solution in harmonic coordinates 
\cite{WEINB}.

To compare our result with the post--Newtonian expansion of the result of Ref.~\cite{Bern:2021dqo}
to 6PN we perform the following canonical transformation
\begin{eqnarray}
  \label{eq:2}
  \bar H  &=& H + \{H , g\} + \frac{1}{2!} \{\{H , g\} , g\} +
  \frac{1}{3!} \{\{\{H , g\} , g\}  , g\} \nonumber \\
  && + \frac{1}{4!} \{\{\{\{H , g\} , g\}  , g\}  , g\}  +
     \frac{1}{5!} \{\{\{\{\{H , g\} , g\}  , g\}  , g\} , g\}
     \nonumber \\
    &&+\frac{1}{6!} \{\{\{\{\{\{H , g\} , g\}  , g\}  , g\} , g\} ,
       g\} \,,
\end{eqnarray}
where $\{\cdot,\cdot\}$ denotes the Lie bracket and $H$ and $\bar{H}$ are the Hamiltonians for which the 
transformation is performed. The corresponding expressions have to be expand to the respective 
post-Newtonian order. In the logarithmic terms the scale $r_0 = e^{-\gamma_E/2}/(2 \sqrt{\pi} \mu_1)$ 
appears. Here $\gamma_E$ denotes the Euler--Mascheroni constant and $\mu_1$ is the rescaled mass scale 
appearing in $G_N$ in $D$ dimensions.

The function $g$ inducing the canonical transformation is in general
given by
\begin{equation}
  \label{eq:3}
  g = p.r \sum_{i=-1}^0 \sum_{j,k,l=0} \sum_{m=0}^1  \, \alpha_{ijklm}\,
  \varepsilon^i \,
  r^{-j} \,  p^{2k} \, (p.n)^{2l} \ln^m (r/r_0) \,,
\end{equation}
with $\nu$-dependent coefficients $\alpha_{ijklm}$.  
Using this ansatz and evaluating Eq.~(\ref{eq:2}) the corresponding
explicit transformation can be found. The generating function, $g_{\rm harm}^{\rm isotr}$,
is given in Appendix~\ref{sec:A}, Eq.~(\ref{cantr1}), mapping our result to that of \cite{Bern:2021dqo}.
The potential contributions to the scattering angle have been already found to be the same up to 5PN,
cf.~\cite{Bern:2021dqo}, referring to the Hamiltonian derived in \cite{Blumlein:2020pyo}.
Here we proved that this applies to the potential contributions to all observables to 6PN.

Next we compare to the part of the local contributions in Ref.~\cite{Bini:2020nsb}, Eq.~(7.29) 
to $O(G_N^4)$ and the lower order terms in $G_N$, which stem from the potential terms. 
These are all contributions with the exception of the purely rational terms of order 
$\nu^1, \nu^2$ and $\nu^3$, which contain also local tail contributions \cite{TAIL,Bini:2020hmy}. We determine 
the generating function $\tilde{g}_{\rm harm}^{\rm EOB}$, given in Appendix~\ref{sec:A}, 
Eq.~(\ref{cantr2}). Again we find full agreement in all these terms, which also will imply agreement for 
the corresponding contributions to the scattering angle.
In this way a thorough test of all results up to 6PN stemming from the potential contributions to 
$O(G_N^4)$ has been be obtained.

\vspace*{4mm}
\noindent
{\bf Acknowledgment.}~
We thank Z.~Bern for providing a computer-readable version of the Hamiltonian obtained 
in~\cite{Bern:2021dqo}. This project has received funding from the European Union's Horizon 2020 research 
and innovation programme under the Marie Sk\l{}odowska--Curie grant agreement No. 764850, SAGEX and KMP Berlin.

\appendix
\section{The generators of the canonical transformations}
\label{sec:A}

\vspace*{1mm}
\noindent
The generator of the canonical transformation from harmonic coordinates to the isotropic coordinates 
used in \cite{Bern:2021dqo} is given by
\begin{eqnarray}
\label{cantr1}
\lefteqn{g^{\rm isotr}_{\rm harm} = }  \nonumber\\ &&
 p.r \Biggl\{
        \frac{1}{r} \Biggl[
                \frac{\nu }{2}
                -\frac{1}{8} p^2 \nu ^2
                +\frac{(p.n)^2 \nu ^2}{8}
                -\frac{1}{16} p^4 \nu ^3
                -\frac{1}{48} p^2 (p.n)^2 \nu ^3
                +\frac{(p.n)^4 \nu ^3}{16}
                +\frac{5 (p.n)^6 \nu ^4}{128}
                \nonumber\\
                &&+\frac{7 (p.n)^8 \nu ^5}{256}
                +\frac{21 (p.n)^{10} \nu ^6}{1024}
                -\frac{5}{384} p^4 (p.n)^2 \nu ^3 (4+\nu )
                -\frac{1}{128} p^2 (p.n)^4 \nu ^3 (4+\nu )
                \nonumber\\
                &&                -\frac{1}{128} p^6 \nu ^3 (-96+5 \nu )
+p^8 \Biggl(
                        \frac{91 \nu ^3}{128}-\frac{97 \nu ^4}{64}-\frac{7 \nu ^5}{256}\Biggr)
                +p^6 (p.n)^2 \Biggl(
                        -\frac{51 \nu ^3}{256}+\frac{337 \nu ^4}{768}-\frac{7 \nu ^5}{768}\Biggr)
\nonumber\\ &&                
+p^4 (p.n)^4 \Biggl(
                        \frac{3 \nu ^3}{256}+\frac{17 \nu ^4}{256}-\frac{7 \nu ^5}{1280}\Biggr)
                +p^2 (p.n)^6 \Biggl(
                        \frac{5 \nu ^3}{256}-\frac{25 \nu ^4}{256}-\frac{\nu ^5}{256}\Biggr)
                +p^{10}
                 \Biggl(
                        \frac{59 \nu ^3}{32}-\frac{4101 \nu ^4}{512}
\nonumber\\ &&
+\frac{2133 \nu ^5}{256}-\frac{21 \nu 
^6}{1024}\Biggr)
                +p^8 (p.n)^2 \Biggl(
                        -\frac{221 \nu ^3}{512}+\frac{115 \nu ^4}{64}-\frac{1091 \nu ^5}{768}-\frac{7 \nu 
^6}{1024}\Biggr)
                +p^6 (p.n)^4 \Biggl(
                        \frac{177 \nu ^3}{2560}
                \nonumber\\ &&
-\frac{111 \nu ^4}{640}-\frac{1319 \nu ^5}{5120}-\frac{21 
\nu ^6}{5120}\Biggr)
                +p^4 (p.n)^6 \Biggl(
                        \frac{11 \nu ^3}{512}-\frac{23 \nu ^4}{128}+\frac{463 \nu ^5}{1024}
-\frac{3 \nu 
^6}{1024}\Biggr)
\nonumber\\ &&                
+p^2 (p.n)^8 \Biggl(
                        -\frac{7 \nu ^3}{512}+\frac{49 \nu ^4}{512}-\frac{49 \nu ^5}{256}-\frac{7 \nu 
^6}{3072}\Biggr)
        \Biggr]
\nonumber\\ &&
        +\frac{1}{r^2} \Biggl[
                \frac{1}{4}
                -\frac{3 \nu }{4}
                +\frac{\nu ^2}{4}
                +p^2 \Biggl(
                        \frac{9 \nu }{4}-\frac{5 \nu ^2}{8}
+\frac{\nu ^3}{16}\Biggr)
+(p.n)^2 \Biggl(
                        -\frac{7 \nu }{12}-\frac{11 \nu ^2}{2}+\frac{5 \nu ^3}{48}\Biggr)
                \nonumber\\
                &&
 +p^2 (p.n)^2 \Biggl(
                        -\frac{79 \nu }{96}+\frac{413 \nu ^2}{32}+\frac{797 \nu ^3}{48}+\frac{\nu 
^4}{48}\Biggr)
\nonumber\\ &&                
+p^4 \Biggl(
                        \frac{309 \nu }{64}-\frac{635 \nu ^2}{64}-
                        \frac{35 \nu ^3}{4}+\frac{\nu ^4}{32}\Biggr)
                +(p.n)^4 \Biggl(
                        -\frac{487 \nu }{960}-\frac{181 \nu ^2}{64}-\frac{142 \nu ^3}{15}+\frac{7 \nu 
^4}{96}\Biggr)
\nonumber\\ &&
                +p^4 (p.n)^2 \Biggl(
                        -\frac{715 \nu }{192}+\frac{31 \nu ^2}{6}+\frac{439 \nu ^3}{48}-\frac{9001 \nu 
^4}{96}+\frac{7 \nu ^5}{768}\Biggr)
+p^2 (p.n)^4 \Biggl(
                        \frac{19 \nu }{12}+\frac{3949 \nu ^2}{960}-\frac{8711 \nu ^3}{960}
\nonumber\\ &&
+\frac{116237 \nu 
^4}{960}+\frac{49 \nu ^5}{3840}\Biggr)
                +p^6 \Biggl(
                        \frac{1689 \nu }{128}-\frac{5601 \nu ^2}{128}+\frac{475 \nu ^3}{16}+\frac{3543 \nu 
^4}{128}+\frac{5 \nu ^5}{256}\Biggr)
                \nonumber\\
                &&+(p.n)^6 \Biggl(
                        -\frac{159 \nu }{224}-\frac{751 \nu ^2}{224}+\frac{2263 \nu ^3}{280}-\frac{89981 
\nu ^4}{2240}+\frac{15 \nu ^5}{256}\Biggr)
                \nonumber\\
                &&+p^4 (p.n)^4 \Biggl(
                        -\frac{5021 \nu }{1920}-\frac{27691 \nu ^2}{3840}+\frac{9845 \nu 
^3}{256}-\frac{127735 \nu ^4}{1536}-\frac{140303 \nu ^5}{7680}+\frac{19 \nu ^6}{3840}\Biggr)
                \nonumber\\
                &&+p^6 (p.n)^2 \Biggl(
                        -\frac{1943 \nu }{384}+
                        \frac{49709 \nu ^2}{768}-\frac{71357 \nu ^3}{384}+\frac{233293 \nu 
^4}{1536}+\frac{83461 \nu ^5}{1536}+\frac{\nu ^6}{192}\Biggr)
                \nonumber\\
                &&+p^2 (p.n)^6 \Biggl(
                        \frac{7783 \nu }{8960}+\frac{11253 \nu ^2}{896}-\frac{75627 \nu 
^3}{2240}+\frac{1258391 \nu ^4}{17920}+\frac{181733 \nu ^5}{17920}+\frac{3 \nu ^6}{320}\Biggr)
                \nonumber\\
                &&+p^8 \Biggl(
                        \frac{14557 \nu }{512}-\frac{90189 \nu ^2}{512}+\frac{21797 \nu ^3}{64}-\frac{97743 
\nu ^4}{512}-\frac{2917 \nu ^5}{256}+\frac{7 \nu ^6}{512}\Biggr)
                \nonumber\\
                &&+(p.n)^8 \Biggl(
                        -\frac{5333 \nu }{11520}-\frac{40343 \nu ^2}{11520}+\frac{226997 \nu 
^3}{20160}-\frac{843847 \nu ^4}{32256}-\frac{76493 \nu ^5}{161280}+\frac{77 \nu ^6}{1536}\Biggr)
        \Biggr]
        \nonumber\\
        &&+\frac{1}{r^3} \Biggl[
                \frac{2789 \nu }{144}
                -\frac{7 \pi ^2 \nu }{8}
                +\frac{5 \nu ^2}{16}
                +\frac{\nu ^3}{16}
                +(p.n)^2 \Biggl(
                        \frac{34973 \nu }{960}-\frac{879 \pi ^2 \nu }{1024}+\frac{151089 \nu 
^2}{1600}
-\frac{69 \pi ^2 \nu ^2}{128}
\nonumber\\ &&
+\frac{239 \nu ^3}{192}+
                        \frac{\nu ^4}{32}\Biggr)
                +p^2 \Biggl(
                        -\frac{824117 \nu }{14400}+\frac{643 \pi ^2 \nu }{1024}+\frac{341089 \nu 
^2}{14400}-\frac{133 \pi ^2 \nu ^2}{64}-\frac{3 \nu ^3}{64}+\frac{\nu ^4}{16}\Biggr)
                \nonumber\\
                &&+p^2 (p.n)^2 \Biggl(
                        -\frac{12459777 \nu }{78400}+\frac{1269 \pi ^2 \nu }{512}-\frac{25111447 \nu 
^2}{470400}-\frac{1455 \pi ^2 \nu ^2}{256}-\frac{6705133 \nu ^3}{14700}
\nonumber\\ &&
+\frac{8109 \pi ^2 \nu 
^3}{512}-\frac{3875 \nu ^4}{192}+\frac{\nu ^5}{64}\Biggr)
                +(p.n)^4 \Biggl(
                        \frac{15911 \nu }{192}-\frac{6015 \pi ^2 \nu }{4096}-\frac{775711 \nu 
^2}{188160}-\frac{2025 \pi ^2 \nu ^2}{4096}
\nonumber\\ &&
+\frac{5196367 \nu ^3}{12544}+\frac{9375 \pi ^2 \nu 
^3}{1024}+\frac{5913 \nu ^4}{640}+\frac{47 \nu ^5}{1280}\Biggr)
                +p^4 \Biggl(
                        \frac{12576721 \nu }{705600}-\frac{5089 \pi ^2 \nu }{4096}+\frac{912076073 \nu 
^2}{2822400}
\nonumber\\ &&
+\frac{15153 \pi ^2 \nu ^2}{4096}+\frac{281619239 \nu ^3}{8467200}-\frac{15387 \pi ^2 \nu 
^3}{1024}+\frac{41 \nu ^4}{4}+\frac{15 \nu ^5}{256}\Biggr)
                +p^4 (p.n)
                ^2 \Biggl(
                        -
                        \frac{159648631 \nu }{4233600}
\nonumber\\ &&
-\frac{398991 \pi ^2 \nu }{131072}+\frac{1978082033 
\nu ^2}{1209600}+\frac{33957 \pi ^2 \nu ^2}{8192}-\frac{55736243399 \nu ^3}{33868800}-\frac{356457 \pi ^2 
\nu ^3}{8192}
\nonumber\\ &&
+\frac{66817696627 \nu ^4}{16934400}+\frac{525735 \pi ^2 \nu ^4}{4096}+\frac{1544101 \nu 
^5}{9216}+\frac{23 \nu ^6}{2304}\Biggr)
                +p^2 (p.n)^4 \Biggl(
                        -\frac{128052787 \nu }{1693440}
\nonumber\\ &&
+\frac{621015 \pi ^2 \nu }{131072}
-\frac{263026853 
\nu ^2}{211680}-\frac{12495 \pi ^2 \nu ^2}{2048}+\frac{1871276137 \nu ^3}{1354752}+\frac{277755 \pi ^2 \nu 
^3}{8192}
\nonumber\\ &&
-\frac{13422417097 \nu ^4}{1693440}-\frac{1405305 \pi ^2 \nu ^4}{8192}-\frac{8877829 \nu 
^5}{46080}+\frac{181 \nu ^6}{11520}\Biggr)
                +(p.n)^6 \Biggl(
                        -\frac{4066361 \nu }{120960}
\nonumber\\ &&
-\frac{325045 \pi ^2 \nu }{131072}+\frac{53428687 \nu 
^2}{103680}-\frac{2625 \pi ^2 \nu ^2}{4096}-\frac{233917561 \nu ^3}{967680}+\frac{87185 \pi ^2 \nu 
^3}{4096}+\frac{1274653087 \nu ^4}{290304}
\nonumber\\ &&
+\frac{1015 \pi ^2 \nu ^4}{32}+\frac{4666447 \nu 
^5}{64512}+\frac{469 \nu ^6}{11520}\Biggr)
                +p^6 \Biggl(
                        \frac{2207017 \nu }{17640}+\frac{161901 \pi ^2 \nu }{131072}-\frac{11876924429 \nu 
^2}{50803200}
\nonumber \\ &&
-\frac{32955 \pi ^2 \nu ^2}{8192}-\frac{3077092201 \nu ^3}{2903040}+\frac{7023 \pi ^2 \nu 
^3}{512}+\frac{1347718537 \nu ^4}{2822400}-\frac{213837 \pi ^2 \nu ^4}{8192}-\frac{53671 \nu 
^5}{1024}
\nonumber\\ &&
+\frac{7 \nu ^6}{128}\Biggr)
        \Biggr]
        +\frac{1}{r^4} \Biggl[
                \frac{1}{32}
                +\frac{7799 \nu }{180}
                +\frac{50725 \pi ^2 \nu }{12288}
                -\frac{62411 \nu ^2}{450}
\nonumber\\ &&                
+\frac{1123 \pi ^2 \nu ^2}{128}
                +\frac{5 \nu ^3}{32}
                -\frac{\nu ^4}{48}
                +p^2 \Biggl(
                        \frac{120067861 \nu }{117600}-\frac{811375 \pi ^2 \nu }{8192}-
                        \frac{5554269127 \nu ^2}{2116800}
\nonumber\\ &&
+\frac{3786257 \pi ^2 \nu 
^2}{24576}+\frac{130885177 \nu ^3}{75600}+\frac{3805 \pi ^2 \nu ^3}{1536}+\frac{\nu ^4}{2}-\frac{\nu 
^5}{96}\Biggr)
+(p.n)^2 \Biggl(
                        -\frac{62677409 \nu }{50400}
                \nonumber\\
                &&
+\frac{4233 \pi ^2 \nu }{32}
+\frac{559728613 \nu 
^2}{211680}-\frac{649303 \pi ^2 \nu ^2}{6144}-\frac{956636399 \nu ^3}{302400}+\frac{34639 \pi ^2 \nu 
^3}{768}-\frac{79 \nu ^4}{64}-\frac{5 \nu ^5}{576}\Biggr)
\nonumber\\ &&                
+p^2 (p.n)^2 \Biggl(
                        -\frac{402835634819 \nu }{19051200}+\frac{24771080291 \pi ^2 \nu 
}{18874368}+\frac{9455970781 \nu ^2}{77760}-\frac{87600553 \pi ^2 \nu ^2}{24576}
\nonumber\\ &&
-\frac{2626094395199 \nu 
^3}{38102400}-\frac{359797489 \pi ^2 \nu ^3}{73728}-\frac{233531135 \nu ^4}{6048}+\frac{52259255 \pi ^2 \nu 
^4}{9216}+\frac{219 \nu ^5}{256}-\frac{11 \nu ^6}{1152}\Biggr)
\nonumber\\ &&                
+p^4 \Biggl(
                        \frac{18919792957 \nu }{2540160}-\frac{3779149295 \pi ^2 \nu 
}{8388608}-\frac{795747395203 \nu ^2}{25401600}+\frac{16564239 \pi ^2 \nu ^2}{32768}+\frac{1916499983 \nu 
^3}{907200}
\nonumber\\ &&                        
+
                        \frac{257894995 \pi ^2 \nu ^3}{98304}-\frac{5258571887 \nu 
^4}{604800}+\frac{5306093 \pi ^2 \nu ^4}{6144}-\frac{303 \nu ^5}{512}-\frac{\nu ^6}{768}\Biggr)
\nonumber\\ &&
                +(p.n)^4 \Biggl(
                        \frac{157481938247 \nu }{12700800}
-\frac{2512372145 \pi ^2 \nu 
}{3145728}-\frac{2514012457207 \nu ^2}{25401600}+\frac{34311647 \pi ^2 \nu ^2}{8192}
\nonumber\\ &&
+\frac{2630001389207 
\nu ^3}{25401600}
-\frac{8715005 \pi ^2 \nu ^3}{12288}
+\frac{525737859 \nu ^4}{9800}-\frac{52792535 \pi ^2 
\nu ^4}{6144}+\frac{1031 \nu ^5}{1536}+\frac{11 \nu ^6}{3840}\Biggr)
        \Biggr]
\nonumber\\ &&
+\frac{1}{\ep} \Biggl[
                \frac{1}{r^3} \Biggl[
                        -\frac{17 \nu }{6}
                        +\frac{1}{90} p^2 \nu  (585+4 \nu )
                        -\frac{1}{3} (p.n)^2 \nu  (12+37 \nu )
                        -(p.n)^4 \Biggl(
                                11 \nu 
-\frac{13 \nu ^2}{6}
\nonumber\\ &&
+\frac{413 \nu ^3}{12}\Biggr)
                        +p^4 \Biggl(
                                \frac{759 \nu }{40}-\frac{69887 \nu ^2}{1260}-\frac{10001 \nu ^3}{720}\Biggr)
                        +p^2 (p.n)^2 \Biggl(
                                \frac{16 \nu }{5}+28 \nu ^2+\frac{43567 \nu ^3}{840}\Biggr)
                        \nonumber\\
                        &&+p^6 \Biggl(
                                \frac{5777 \nu }{140}-\frac{780211 \nu ^2}{4320}+\frac{1631395 \nu 
^3}{6048}-\frac{2975597 \nu ^4}{30240}\Biggr)
+p^2 (p.n)^4 \Biggl(
                                \frac{213 \nu }{14}+\frac{6851 \nu ^2}{84}
\nonumber\\ &&
+
                                \frac{2591 \nu ^3}{224}
-\frac{12837 \nu ^4}{224}\Biggr)
                        +p^4 (p.n)^2 \Biggl(
                                -\frac{1801 \nu }{280}-\frac{5359 \nu ^2}{210}-\frac{17339 \nu 
^3}{240}+\frac{23201 \nu ^4}{1120}\Biggr)
                        \nonumber\\
                        &&+(p.n)^6 \Biggl(
                                -\frac{37 \nu }{2}-\frac{106 \nu ^2}{9}-\frac{290 \nu ^3}{9}+\frac{1789 \nu 
^4}{18}\Biggr)
                \Biggr]
                +\frac{1}{r^4} \Biggl[
                        -\frac{59 \nu }{30}
                        -\frac{443 \nu ^2}{60}
                        +p^2 \Biggl(
                                -\frac{5347 \nu }{168}
\nonumber\\ &&
+\frac{4828 \nu ^2}{35}-\frac{15629 \nu ^3}{126}\Biggr)
                        +(p.n)^2 \Biggl(
                                \frac{4087 \nu }{140}-\frac{35153 \nu ^2}{252}+\frac{9067 \nu ^3}{42}\Biggr)
                        +p^2 (p.n)^2 \Biggl(
                                \frac{2276941 \nu }{3780}
\nonumber\\ &&
-\frac{13614733 \nu ^2}{1890}+\frac{10245317 \nu 
^3}{1080}-\frac{2189263 \nu ^4}{1008}\Biggr)
                        +p^4 \Biggl(
                                -\frac{2461261 \nu }{10080}+\frac{23907 \nu ^2}{10}
\nonumber\\ &&
-\frac{1833827 \nu 
^3}{1260}-\frac{846893 \nu ^4}{1680}\Biggr)
                        +(p.n)^4 \Biggl(
                                -\frac{258989 \nu }{1260}+\frac{10783751 \nu ^2}{2520}-\frac{38512301 \nu 
^3}{5040}
\nonumber\\ &&
+
                                \frac{616781 \nu ^4}{240}\Biggr)
                \Biggr]
        \Biggr]
        +\ln \left(
                \frac{r}{r_0}
        \right)
\Biggl[
\frac{1}{r^3} \Biggl[
                        -17 \nu 
                        +\frac{1}{15} p^2 \nu  (585+4 \nu )
                        -2 (p.n)^2 \nu  (12+37 \nu )
\nonumber\\ &&                        
-\frac{1}{2} (p.n)^4 \nu  \Biggl(
                                132-26 \nu +413 \nu ^2\Biggr)
                        +p^4 \Biggl(
                                \frac{2277 \nu }{20}-\frac{69887 \nu ^2}{210}-\frac{10001 \nu ^3}{120}\Biggr)
                        \nonumber\\
                        &&+p^2 (p.n)^2 \Biggl(
                                \frac{96 \nu }{5}+168 \nu ^2+\frac{43567 \nu ^3}{140}\Biggr)
                        +p^6 \Biggl(
                                \frac{17331 \nu }{70}-\frac{780211 \nu ^2}{720}+\frac{1631395 \nu 
^3}{1008}
\nonumber\\ &&
-\frac{2975597 \nu ^4}{5040}\Biggr)
                        +p^2 (p.n)^4 \Biggl(
                                \frac{639 \nu }{7}+\frac{6851 \nu ^2}{14}+\frac{7773 \nu 
^3}{112}-\frac{38511 \nu ^4}{112}\Biggr)
                        +p^4 (p.n)^2 \Biggl(
                                -\frac{5403 \nu }{140}
\nonumber\\ &&
-\frac{5359 \nu ^2}{35}-\frac{17339 \nu 
^3}{40}+\frac{69603 \nu ^4}{560}\Biggr)
                        +(p.n)^6 \Biggl(
                                -111 \nu -\frac{212 \nu ^2}{3}-\frac{580 \nu ^3}{3}+\frac{1789 \nu 
^4}{3}\Biggr)
                \Biggr]
\nonumber\\ &&
                +\frac{1}{r^4} \Biggl(
                        -\frac{236 \nu }{15}
                        -\frac{619 \nu ^2}{10}
                        +p^2 \Biggl(
                                -\frac{5347 \nu }{21}+\frac{39079 \nu ^2}{35}-\frac{34728 \nu ^3}{35}\Biggr)
                        +(p.n)^2 \Biggl(
                                \frac{8174 \nu }{35}
\nonumber\\ &&
-\frac{141935 \nu ^2}{126}
+\frac{540121 \nu 
^3}{315}\Biggr)
                        +p^2 (p.n)^2 \Biggl(
                                \frac{4553882 \nu }{945}-\frac{217954987 \nu ^2}{3780}+\frac{71879222 \nu 
^3}{945}
                        \nonumber\\
                        &&	-\frac{6209047 \nu ^4}{360}\Biggr)
+p^4 \Biggl(
                                -\frac{2461261 \nu }{1260}+\frac{767301 \nu ^2}{40}-\frac{29764649 \nu 
^3}{2520}-\frac{10268647 \nu ^4}{2520}\Biggr)
                        \nonumber\\
                        &&+(p.n)^4 \Biggl(
                                -\frac{517978 \nu }{315}+\frac{10779278 \nu ^2}{315}-\frac{154121549 \nu 
^3}{2520}+\frac{17201353 \nu ^4}{840}\Biggr)
                \Biggr]
        \Biggr]
\Biggr\}. 
\end{eqnarray} 

The generator of the canonical transformation to the 6PN EOB Hamiltonian reads
\begin{align}
\label{cantr2}
& g_{\rm harm}^{\rm EOB} =\notag\\
{}&p.r\*\Biggl\{\frac{p^{2}\*\nu}{2}-\frac{p^{4}\*\nu}{8}+\frac{5\*p^{10}\*\nu^{3}}{96}+p^{6}\*\biggl(\frac{\nu}{16}-\frac{\nu^{2}}{16}\bigg)+p^{8}\*\biggl(\frac{5\*\nu^{2}}{64}-\frac{\nu^{3}}{48}\biggr)+p^{12}\*\biggl(\frac{\nu^{4}}{96}+\frac{\nu^{5}}{240}\biggr)\notag\\
&   +\frac{1}{r^2}\*\biggl\{\frac{5\*\nu}{4}-\frac{\nu^{2}}{4}+p^{8}\*\biggl(-\frac{62541\*\nu^{4}}{256}-\frac{102821\*\nu^{5}}{1536}+\frac{\nu^{6}}{192}\biggr)+\biggl(\frac{4\*\nu}{3}-\frac{83\*\nu^{2}}{24}-\frac{\nu^{3}}{8}\biggr)\*(p.n)^{2}\notag\\
&      +\biggl(\frac{1333\*\nu^{2}}{960}+\frac{31\*\nu^{3}}{80}+\frac{\nu^{4}}{48}\biggr)\*(p.n)^{4}+\biggl(\frac{21565\*\nu^{3}}{1344}-\frac{63677\*\nu^{4}}{2688}+\frac{13\*\nu^{5}}{128}\biggr)\*(p.n)^{6}\notag\\
&+\biggl(-\frac{2351599\*\nu^{4}}{32256}         +\frac{3989339\*\nu^{5}}{32256}-\frac{9\*\nu^{6}}{1280}\biggr)\*(p.n)^{8}+p^{6}\*\biggl[\frac{16783\*\nu^{3}}{384}+\frac{14675\*\nu^{4}}{384}-\frac{\nu^{5}}{384}\notag\\
&            +\biggl(\frac{54537\*\nu^{4}}{512}+\frac{113807\*\nu^{5}}{1440}-\frac{23\*\nu^{6}}{11520}\biggr)\*(p.n)^{2}\biggr]+p^{4}\*\biggl[-\frac{1681\*\nu^{2}}{192}-\frac{143\*\nu^{3}}{16}-\frac{\nu^{4}}{24}\notag\\
&         +\biggl(\frac{6157\*\nu^{3}}{384}-\frac{100025\*\nu^{4}}{1152}+\frac{13\*\nu^{5}}{576}\biggr)\*(p.n)^{2}+\biggl(-\frac{804613\*\nu^{4}}{7680}+\frac{1252559\*\nu^{5}}{5760}+\frac{41\*\nu^{6}}{1152}\biggr)\*(p.n)^{4}\biggr]\notag\\
&         +p^{2}\*\biggl[\frac{29\*\nu}{6}-\frac{19\*\nu^{2}}{24}+\frac{5\*\nu^{3}}{24}+\biggl(\frac{817\*\nu^{2}}{96}+\frac{493\*\nu^{3}}{32}+\frac{\nu^{4}}{48}\biggr)\*(p.n)^{2}+\biggl(-\frac{793\*\nu^{3}}{20}+\frac{484729\*\nu^{4}}{5760}\notag\\
&         -\frac{199\*\nu^{5}}{1440}\biggr)\*(p.n)^{4}+\biggl(\frac{7679449\*\nu^{4}}{53760}-\frac{481625\*\nu^{5}}{1344}-\frac{61\*\nu^{6}}{1920}\biggr)\*(p.n)^{6}\biggr]\biggr\}+\frac{1}{r}\*\biggl\{-1-\frac{\nu}{2}\notag\\
&      +p^{10}\*\biggl(-\frac{274391\*\nu^{4}}{30720}+\frac{118049\*\nu^{5}}{10240}\biggr)+\biggl(-\frac{\nu}{2}-\frac{\nu^{2}}{8}\biggr)\*(p.n)^{2}-\frac{\nu^{2}\*(p.n)^{4}}{4}+\frac{5\*\nu^{4}\*(p.n)^{6}}{128}+\frac{7\*\nu^{4}\*(p.n)^{8}}{48}\notag\\
&         -\frac{63\*\nu^{6}\*(p.n)^{10}}{1024}+p^{8}\*\biggl[\frac{569\*\nu^{3}}{512}-\frac{1927\*\nu^{4}}{768}+\frac{\nu^{5}}{768}+\biggl(\frac{14231\*\nu^{4}}{10240}-\frac{64907\*\nu^{5}}{92160}-\frac{\nu^{6}}{7680}\biggr)\*(p.n)^{2}\biggr]\notag\\
&         +p^{6}\*\biggl[\frac{103\*\nu^{2}}{192}+\frac{229\*\nu^{3}}{384}+\biggl(-\frac{205\*\nu^{3}}{384}+\frac{145\*\nu^{4}}{768}+\frac{\nu^{5}}{4608}\biggr)\*(p.n)^{2}+\biggl(\frac{629\*\nu^{4}}{5120}-\frac{20833\*\nu^{5}}{30720}\notag\\
&         +\frac{19\*\nu^{6}}{10240}\biggr)\*(p.n)^{4}\biggr]+p^{4}\*\biggl[\frac{7\*\nu}{16}-\frac{37\*\nu^{2}}{96}-\frac{\nu^{3}}{48}+\biggl(-\frac{7\*\nu^{2}}{96}+\frac{61\*\nu^{3}}{384}+\frac{\nu^{4}}{96}\biggr)\*(p.n)^{2}+\biggl(\frac{65\*\nu^{3}}{256}+\frac{29\*\nu^{4}}{128}\notag\\
&         +\frac{37\*\nu^{5}}{7680}\biggr)\*(p.n)^{4}+\biggl(\frac{191\*\nu^{4}}{1024}+\frac{241\*\nu^{5}}{6144}-\frac{265\*\nu^{6}}{6144}\biggr)\*(p.n)^{6}\biggr]+p^{2}\*\biggl[-\frac{5\*\nu}{4}+\frac{\nu^{2}}{4}+\biggl(\frac{\nu}{4}+\frac{7\*\nu^{2}}{24}\notag\\
&         +\frac{\nu^{3}}{96}\biggr)\*(p.n)^{2}+\biggl(\frac{\nu^{2}}{4}-\frac{\nu^{3}}{16}-\frac{7\*\nu^{4}}{128}\biggr)\*(p.n)^{4}+\biggl(\frac{5\*\nu^{3}}{256}-\frac{203\*\nu^{4}}{768}-\frac{7\*\nu^{5}}{1536}\biggr)\*(p.n)^{6}+\biggl(-\frac{203\*\nu^{4}}{768}\notag\\
&            +\frac{35\*\nu^{5}}{384}+\frac{161\*\nu^{6}}{1536}\biggr)\*(p.n)^{8}\biggr]\biggr\}+\frac{1}{
\varepsilon}\*\biggl\{\frac{1}{r^4}\*\biggl[-\frac{613\*\nu^{2}}{60}-\frac{12805747\*p^{4}\*\nu^{4}}{30240}+\frac{112745\*\nu^{3}\*(p.n)^{2}}{504}\notag\\
&         +\frac{2677891\*\nu^{4}\*(p.n)^{4}}{1440}+p^{2}\*\biggl(-\frac{16671\*\nu^{3}}{140}-\frac{27275749\*\nu^{4}\*(p.n)^{2}}{15120}\biggr)\biggr]+\frac{1}{r^3}\*\biggl[-\frac{17\*\nu}{6}-\frac{4900249\*p^{6}\*\nu^{4}}{60480}\notag\\
&         -\frac{97\*\nu^{2}\*(p.n)^{2}}{12}-\frac{57\*\nu^{3}\*(p.n)^{4}}{8}+\frac{1471\*\nu^{4}\*(p.n)^{6}}{8}+p^{4}\*\biggl(-\frac{4061\*\nu^{3}}{288}+\frac{32327\*\nu^{4}\*(p.n)^{2}}{1120}\biggr)\notag\\
&         +p^{2}\*\biggl(\frac{271\*\nu^{2}}{360}+\frac{40921\*\nu^{3}\*(p.n)^{2}}{840}-\frac{45349\*\nu^{4}\*(p.n)^{4}}{224}\biggr)\biggr]\biggr\}+\frac{1}{r^4}\*\Biggl[\frac{1187\*\pi^{2}\*\nu}{1024}+\frac{7\*\nu^{3}}{96}-\frac{\nu^{4}}{6}\notag\\
&      +p^{4}\*\biggl\{-\frac{59689843\*\pi^{2}\*\nu}{131072}+\frac{14986753\*\pi^{2}\*\nu^{2}}{24576}+\frac{14964907\*\pi^{2}\*\nu^{3}}{6144}-\frac{77051\*\nu^{5}}{7680}-\frac{139\*\nu^{6}}{1920}\notag\\
&         +\nu^{4}\*\biggl[-\frac{60034957159}{5644800}+\frac{5048789\*\pi^{2}}{6144}-\frac{5621293}{1680}\*\ln\biggl(\frac{r}{r_{0}}\biggr)\biggr]\biggr\}+p^{2}\*\Biggl(-\frac{624073\*\pi^{2}\*\nu}{6144}\notag\\
&         +\frac{920315\*\pi^{2}\*\nu^{2}}{6144}-\frac{7\*\nu^{4}}{40}+\frac{107\*\nu^{5}}{480}+(p.n)^{2}\*\biggl\{\frac{1555146239\*\pi^{2}\*\nu}{1179648}-\frac{6894431\*\pi^{2}\*\nu^{2}}{2048}-\frac{47430287\*\pi^{2}\*\nu^{3}}{9216}\notag\\
&            +\frac{174347\*\nu^{5}}{11520}+\frac{17\*\nu^{6}}{384}+\nu^{4}\*\biggl[-\frac{73238156053}{1693440}+\frac{52850153\*\pi^{2}}{9216}-\frac{7333009}{504}\*\ln\biggl(\frac{r}{r_{0}}\biggr)\biggr]\biggr\}\notag\\
&         +\nu^{3}\*\biggl[\frac{183707899}{100800}-\frac{185\*\pi^{2}}{32}-\frac{200269}{210}\*\ln\biggl(\frac{r}{r_{0}}\biggr)\biggr]\Biggr)+\nu^{2}\*\biggl[-\frac{902701}{7200}+\frac{263\*\pi^{2}}{32}-\frac{789}{10}\*\ln\biggl(\frac{r}{r_{0}}\biggr)\biggr]\notag\\
&         +(p.n)^{2}\*\biggl\{\frac{2317781\*\pi^{2}\*\nu}{18432}-\frac{89525\*\pi^{2}\*\nu^{2}}{768}+\frac{3829\*\nu^{4}}{2880}-\frac{377\*\nu^{5}}{2880}+\nu^{3}\*\biggl[-\frac{32236479}{11200}+\frac{23879\*\pi^{2}}{768}\notag\\
&         +\frac{2263951}{1260}\*\ln\biggl(\frac{r}{r_{0}}\biggr)\biggr]\biggr\}+(p.n)^{4}\*\biggl\{-\frac{1581642941\*\pi^{2}\*\nu}{1966080}+\frac{464633167\*\pi^{2}\*\nu^{2}}{122880}-\frac{1145483\*\pi^{2}\*\nu^{3}}{3072}\notag\\
&      -\frac{167053\*\nu^{5}}{23040}+\frac{1403\*\nu^{6}}{11520}+\nu^{4}\*\biggl[\frac{7419092299}{115200}-\frac{53579429\*\pi^{2}}{6144}+\frac{25070309}{1680}\*\ln\biggl(\frac{r}{r_{0}}\biggr)\biggr]\biggr\}\Biggr]\notag\\
&         +\frac{1}{r^3}\*\Biggl[-\frac{3\*\nu^{2}}{16}-\frac{3\*\nu^{3}}{16}+p^{6}\*\biggl\{\frac{161901\*\pi^{2}\*\nu}{131072}-\frac{20583\*\pi^{2}\*\nu^{2}}{8192}+\frac{226655\*\pi^{2}\*\nu^{3}}{24576}+\frac{981061\*\nu^{5}}{15360}-\frac{3\*\nu^{6}}{256}\notag\\
&      +\nu^{4}\*\biggl[\frac{22497180479}{33868800}-\frac{65287\*\pi^{2}}{8192}-\frac{4900249}{10080}\*\ln\biggl(\frac{r}{r_{0}}\biggr)\biggr]\biggr\}+(p.n)^{2}\*\biggl\{-\frac{879\*\pi^{2}\*\nu}{1024}-\frac{15\*\nu^{3}}{64}-\frac{\nu^{4}}{8}\notag\\
&+\nu^{2}\*\biggl[\frac{109921}{2400}+\frac{99\*\pi^{2}}{128}            -\frac{97}{2}\*\ln\biggl(\frac{r}{r_{0}}\biggr)\biggr]\biggr\}+(p.n)^{4}\*\biggl\{-\frac{6015\*\pi^{2}\*\nu}{4096}+\frac{6765\*\pi^{2}\*\nu^{2}}{4096}-\frac{7073\*\nu^{4}}{1440}\notag\\
&+\frac{697\*\nu^{5}}{11520}+\nu^{3}\*\biggl[\frac{67938979}{564480}+\frac{9635\*\pi^{2}}{1024}-\frac{171}{4}\*\ln\biggl(\frac{r}{r_{0}}\biggr)\biggr]\biggr\}+\nu\*\biggl[\frac{1795}{72}-\frac{7\*\pi^{2}}{8}-17\*\ln\biggl(\frac{r}{r_{0}}\biggr)\biggr]\notag\\
&            +(p.n)^{6}\*\biggl\{-\frac{325045\*\pi^{2}\*\nu}{131072}+\frac{36855\*\pi^{2}\*\nu^{2}}{8192}+\frac{168035\*\pi^{2}\*\nu^{3}}{8192}-\frac{17410529\*\nu^{5}}{322560}+\frac{1489\*\nu^{6}}{7680}\notag\\
&         +\nu^{4}\*\biggl[\frac{1289317591}{414720}-\frac{3885\*\pi^{2}}{2048}+\frac{4413}{4}\*\ln\biggl(\frac{r}{r_{0}}\biggr)\biggr]\biggr\}+p^{4}\*\Biggl(-\frac{5089\*\pi^{2}\*\nu}{4096}+\frac{1681\*\pi^{2}\*\nu^{2}}{512}-\frac{7121\*\nu^{4}}{640}\notag\\
&      -\frac{77\*\nu^{5}}{1280}+\nu^{3}\*\biggl[-\frac{52415933}{1693440}-\frac{13847\*\pi^{2}}{1024}-\frac{4061}{48}\*\ln\biggl(\frac{r}{r_{0}}\biggr)\biggr]+(p.n)^{2}\*\biggl\{-\frac{398991\*\pi^{2}\*\nu}{131072}\notag\\
&            +\frac{110229\*\pi^{2}\*\nu^{2}}{16384}-\frac{806643\*\pi^{2}\*\nu^{3}}{16384}-\frac{7441013\*\nu^{5}}{46080}+\frac{287\*\nu^{6}}{7680}+\nu^{4}\*\biggl[\frac{17537219339}{4838400}+\frac{597629\*\pi^{2}}{4096}\notag\\
&            +\frac{96981}{560}\*\ln\biggl(\frac{r}{r_{0}}\biggr)\biggr]\biggr\}\Biggr)+p^{2}\*\Biggl(\frac{643\*\pi^{2}\*\nu}{1024}-\frac{15\*\nu^{3}}{64}+\frac{5\*\nu^{4}}{24}+(p.n)^{4}\*\biggl\{\frac{621015\*\pi^{2}\*\nu}{131072}-\frac{237105\*\pi^{2}\*\nu^{2}}{16384}\notag\\
&         +\frac{813695\*\pi^{2}\*\nu^{3}}{16384}+\frac{7072291\*\nu^{5}}{46080}-\frac{2017\*\nu^{6}}{7680}+\nu^{4}\*\biggl[-\frac{41208333631}{6773760}-\frac{1691015\*\pi^{2}}{8192}\notag\\
&            -\frac{136047}{112}\*\ln\biggl(\frac{r}{r_{0}}\biggr)\biggr]\biggr\}+\nu^{2}\*\biggl[\frac{48191}{3600}-\frac{119\*\pi^{2}}{64}+\frac{271}{60}\*\ln\biggl(\frac{r}{r_{0}}\biggr)\biggr]+(p.n)^{2}\*\biggl\{\frac{1269\*\pi^{2}\*\nu}{512}\notag\\
&         
-\frac{30705\*\pi^{2}\*\nu^{2}}{4096}+\frac{105479\*\nu^{4}}{5760}+\frac{23\*\nu^{5}}{720}+\nu^{3}\*\biggl[-\frac{279857647}{705600}+\frac{2409\*\pi^{2}}{128}+\frac{40921}{140}\*\ln\biggl(\frac{r}{r_{0}}\biggr)\biggr]\biggr\}\Biggr)\Biggr]\Biggr\}.
\end{align}
Here we excluded the purely rational terms at orders $\nu, \nu^2$ and $\nu^3$, to which local
parts of the tail terms contribute.

{\footnotesize

}
\end{document}